# Robert Grosseteste's colours

**Amelia Carolina Sparavigna**
**Department of Applied Science and Technology, Politecnico di Torino, Italy**

*Here I am proposing a translation and discussion of the De Colore, one of the short scientific treatises written by Robert Grosseteste. In this very short treatise of the mid-1220s, Grosseteste continued the discussion on light and colours he started in the De Iride. He describes two manners of counting colours: one gives an infinity of tones, the other counts seven colours. In both cases, colours are created by the purity or impurity of the transparent medium when light is passing through it. This medieval framework survived until Newton's experiments with prisms.*

In a recent paper [1], I proposed a translation and discussion of the De Iride, On the rainbow, a treatise written by the medieval scholar Robert Grosseteste [2], probably in the early XIII Century. De Iride can be subdivided in two parts: the first is discussing reflection, refraction and the law of refraction. The second part is devoted to the rainbow and has a short discussion of colours. In the mid-1220s, Grosseteste probably composed another very short treatise, "De Colore", On colours, where he further discusses the generation of colours from light and matter. Before reading the "De Colore", let us shortly remember was Grosseteste proposed in the De Iride. He supposed that colours are coming from the interaction of light with more or less transparent media: color sit lumen admixtum cum diaphano, he wrote. He continued enumerating the features of transparent media and light: a medium is diversified according to its purity and impurity, but the light is fourfold divided. The light is to be divided according to the brightness (claritas), and, of course, obscurity, and according to intensity (multitudo) and tenuity. From these features, the colours arise: secundum harum sex differentiarum connumerationes sint omnium colorum generationes et diversitates, varietas coloris in diversis partibus unius et eiusdem iridis maxime accidit propter multitudinem et paucitatem radiorum solis, that is, according to the six different enumerations [3], the variety of all the colors is generated, colors that appear in the different parts of a rainbow, mainly due to the intensity or tenuity of the rays of sun.

In the "De Iride", Grosseteste does not tell us what the colours are, that is, he does not write down a list of colours. Only one is mentioned, the Hyacinthus: ubi enim est maior radiorum multiplicatio, apparet color magis clarus et luminosus; ubi vero minor est radiorum multiplicatio, apparet color magis attinens hyacinthine et obscuro. Where there is a greater intensity of light, it appears that the colors are more luminous and bright: but where there is less intensity of light, it appears that the color turns to the dark color of Hyacinthus.

The hyacinthos is the name of a blue cornflower [4]. We know by its description in the Natural History of Pliny the Elder that, for the Latin world, the hyacinthos was a precious stone too. Pliny tells that it was very different from amethysts. The great difference between amethyst and hyacinthos is that "the brilliant violet which is so refulgent in the amethystos, is diluted in the other stone." [5] John Bostock, who translated the Pliny's History, reports that this could be the name of the oriental amethyst or violet sapphire, not the hyacinthine of the modern English, meaning a stone of a yellowish colour, as in yellow zircon.

The colour then described by Grosseteste was a violet one.

As previously told, in the mid-1220s, Grosseteste composed the De Colore, where he is discussing the generation of colours too. Here let us report the Latin text, as given in Ref.6. The text is subdivided in several sections. A translation is given and some notes, in the same manner as made for the De iride and the De Generatione Sonorum [7,8].

INC: Color est lux incorporate perspicuo.
EXPL: et sic per artificium omnes modos colorum, quos voluerint visibiliter ostendere possunt. Explicit tractatus de colore secundum Lincolniensem.

*INCIPIT: Color is light incorporated in a transparent medium.*
*END: and so through the skill of combining all the modes, the colours that we want can be rendered visible. This is the end of the treatise on colours by a Lincolnian.*

1 – Color est lux incorporate perspicuo. Perspicui vero duae sunt differentiae: est enim perspicuum aut purum separatum a terrestreitate, aut impurum terrestreitatis admixtione. Lux autem quadrifarie partitur: quia aut est lux clara vel obscura, pauca vel multa. Nec dico lucem multam per subiectum magnum diffusam. Sed in puncto colligitur lux multa, cum speculum concavum opponitur soli et lux cadens super totam superficiem speculi in centrum sphaerae speculi reflectitur. Cuius etiam lucis virtute in ipso centro collecta combustibile citissime inflammatur.

*Color is light incorporated in a transparent medium. But, in fact, there are two different media: there are pure transparent media separated from earth materials or impure media mixed with them. The light, however, is four-fold differentiated: there is the bright and the obscure light, and the intense or the tenuous light. I do not say that an intense light is a light diffused by a great object, but it is the light that we can observe in a point where a large amount of it is collected by means of a concave mirror, and the light falling on the entire surface of the mirror facing the sun is reflected in the center of the sphere of the mirror. And then the power of light collected in this center ignites immediately a combustible material.*

First of all, we find that concave mirrors and the focusing of rays were well known. Let us tell also that Grosseteste's "lux multa" is the intensity of light, because he is clearly referring to the focus of the mirror, where we can have a "radiorum multiplicatio", because in the focus we collect several rays of light.
Grosseteste is distinguishing pure and impure transparent media, impure because of the mixture with earth materials. "Earth" is one of the four classical elements in ancient Greek philosophy and science, commonly associated with qualities of heaviness and matter. Here we can associate the earth materials to the defects and the impurities that we have in transparent media such as crystals. However, since Grosseteste is discussing about colours, we could also suppose that he was trying to merge the pure colours that we can see in the dispersion by a prism to those that we can obtain using pigments. We have for instance the earth pigments for painting, some of them well known since prehistoric times. The clay earth pigments, ochre and sienna for instance, are naturally occurring minerals, principally iron oxides.
For what concerns the dispersion of light from prisms, this was a phenomenon described even in the Natural History by Pliny [9]. Since this book was one of the ancient books that the scholars of the Middle Age used for their treatises on Nature, it is quite probable that Grosseteste experimented with prisms, even if he did not discuss the phenomenon in his treatises.

2 – Lux igitur clara multa in perspicuo puro albedo est. Lux pauca in perspicuo impuro nigredo est. Et in hoc sermone explanatus est sermo Aristotelis et Averrois, qui ponunt nigredinem privationem et albedinem habitum sive formam.

*So an intense bright light in a transparent pure medium is the white. Tenuous light in an impure medium is black. And by telling this, we are explaining what Aristotle and Averroes*

*told, who consider that blackness is lack of light and whiteness is richness or appearance if light.*

"Et hic intendebat Philosophus per 'nigredinem' privationem albedinis", wrote Henricus Bate, a Flemish astronomer around 1300 [10], in his Speculum divinorum et quorundam naturalium [11]. Of course, we could imagine a symbolic meaning too [12]. In the early Middle Ages, Ref.12 tells that black was commonly associated with darkness and evil. Clothes of black colour were worn by Benedictine monks as a sign of humility and penitence. In the 12th century a theological dispute about the colours of clothes broke out between the Cistercian monks, who wore white, and the Benedictines. Pierre the Venerable, a Benedictine, accused the Cistercians of excessive pride in wearing white, but Saint Bernard of Clairvaux, founder of the Cistercians, replied that black was the color of the devil, while white represented purity and "all the virtues" [12,13].

Let us also note that albedo is one of the four major stages of the "magnum opus" of the alchemy; along with nigredo, citrinitas and rubedo. From the nigredo stage, the alchemist undertakes a purification in albedo, passing through a chromatic sequence [14].

3 – Sequitur etiam ex hoc sermone, quod colores proximi albedini, in quibus potest fieri recessus ab albedine et permutatio, septem sunt, nec plures nec pauciores. Similiter septem erunt proximi nigredini, quibus a nigredine versus albedinem ascenditur, donec fiat concursus aliorum septem colorum, quibus ab albedine descenditur. Cum enim albedinis essentiam tria constituant, scilicet lucis multitudo, eiusdemque claritas et perspicui puritas, duobus manentibus cuiuslibet trium potest fieri remissio, eritque per hunc modum trium colorum generatio; vel quolibet trium solo manente, duorum reliquorum erit remissio, et sic fiet aliorum colorum a tribus prioribus trina generatio; aut omnium trium simul erit remissio; et sic in universo ab albedine erit septem colorum immediata progressio.

*It also follows from this speech, that the colors near the white, in which we can arrive regressing from the whiteness and by variation, are seven, no more, no fewer. Similarly, there will be seven colours near the black, which we find when progressing from darkness towards whiteness, until we have a combination with the other seven colors, to which we arrive descending from white. Since the essence of whiteness is made of three features, the intensity of light, its brightness and the purity of the transparent medium, two of them remaining fixed as we like, the third can be relaxed, and then this is how three colours are created, or anyone of these three features remains fixed, and the other two relaxed, and so will have other three colors, besides the three colors of the first triple generation; or all the three features are at the same time relaxed, and so the overall seven colors from the whiteness will directly obtained.*

Let us consider the white as given by brightness, intensity and purity: we can relax one, two or all these three features to obtain seven colours, as in the calculus of combinations shown in the Figure 1.

4 – Consimilis est ratio, per quam ostenditur a nigredine per septem colores illi proximos versus albedinem ascensio. Erunt ergo in universo colores sedecim: duo scilicet extremi et hinc inde septem extremis annexi hinc per intensionem ascendentes illinc per remissionem descendentes ac in medio in idem concurrentes. In quolibet autem colorum mediorum gradus intensionis et remissionis sunt infiniti. Unde qui per numerationem et combinatione eorum, quae intenduntur et remittuntur, multitudinis scilicet et claritatis luminis et etiam puritatis perspicui et oppositorum his, fiunt colores novem, per numerationem graduum intensionis et remissionis erunt infiniti. Quod autem secundum dictum modum se habeat colorum essentia et eorundem multitudo, non solum ratione, verum etiam experimento manifestum est his, qui scientiae naturalis et perspectivae profundius et interius noverunt principia. Quod est, quia sciunt figurare perspicuum, sive fuerit purum sive impurum ita, ut in ipso recipiant lumen clarum, sive si maluit obscurum et per figuram formatam in ipso perspicuo lumen paucum faciant, aut ipsam pro libito multiplicent; et sic per artificium omnes modos colorum, quos voluerint visibiliter ostendere possunt. Explicit tractatus de colore secundum Lincolniensem.

*A similar procedure exists, by means of which we can show that through seven colors from the blackness we can progress towards the white. It would be then in general sixteen combinations: two of course are the ends (white and black), and seven at one end, attached to it by the tension of ascending and at the other end by the remission of descending, merging in the same colours in the middle. Now, in any of the intermediate colours, the grades of tension and remission are infinite. Hence, by counting and combining the features, which are to be intensified and released, that is, the intensity, the brightness of light and the purity of transparent media and their opposites, the colours that can be made are nine; by counting the degrees of tension and remission will be an infinite number of tones of intensity. Now then, it is clear to whom who knew deeply and inwardly the principles of the natural science and of optics, not only by reasoning, but also by experience, that we have, according to the manner stated above, the essence of colours and their multitude. That is, knowing how to form a transparent medium, whether it were pure or impure, in such a way to receive a bright light, or, if it is preferred a dark light, and through the devised form in this very transparent medium, the light is reduced, or multiplied at pleasure; and so through the skill of combining all the modes, the colours that we want can be rendered visible. This is the end of the treatise on colours by a Lincolnian.*

The procedure shown in the Figure 1 from white, can be used to ascending from black to obtain seven colours. At Ref.15, there is a quite interesting image related to the Grosseteste's theory, which is adapted in the Figure 2. We see that, from white we have seven colours, and seven colours from black. But these colours move on two cones, merging in the middle, where we have seven "average" colours. White, black and these seven colours give the nine colours mentioned by Grosseteste. However, he counted the colours in two manners: the first is that based on combinations (Figure 1), the second is based on a continuous scale of tones, as we can have in the palette of the Paint software. As shown in the Figure 3 we have an infinite number of greens: in any case, it is the green.
Let me note that recently, a discussion of the Grosseteste's colours in the RGB space has been proposed in Ref.16. In the paper, the authors are arguing that the colour space described by Grosseteste is explicitly three-dimensional.
For what concerns the Latin text, let us note that Grosseteste is using to describe the light the terms lux and lumen. In Ref.17, it is proposed that lux is light in its source, whereas lumen is reflected or radiated light.
In the Figure 2, adapted from [15], we see a circle of colours obtained considering the average of the colours coming from white and black. Sir Isaac Newton proposed a circle of colours containing seven colours too. He called them Aureus, Flavus, Viridis, Caeruleus, Indicus, Violaceus and Rubeus. Newton used seven colours by analogy to the number of notes in a musical scale [18]. It would be interesting a comparison of Newton's colours with those of Grosseteste. However, Grosseteste did not provide the name of them, besides one, the Hyacinthus which is a violaceus (violet) color.

We can ask ourselves whether the Grosseteste's work had some influences until the Newton's times or not. It seems that this is so, as we can find in the book on the life of Sir Isaac Newton, written by David Brewster [19]. Let us consider what Brewster is telling on colours.

He reports that Newton's friend and tutor, Isaac Barrow (1630-1677), delivered some optical lectures, which were published in 1669. "In the preface of this work – Brewster writes – he acknowledges his obligations to his colleague, Mr. Isaac Newton, for having revised the manuscripts, and corrected several oversights, and made some important suggestions. In the twelfth lecture there are some observations on the nature and origin of colours … According to Dr. Barrow, White is that which discharges a copious light equally clear in every direction; Black is that which does not emit light at all, or which does it very sparingly. Red is that which emits a light more clear than usual, but interrupted by shady interstices. Blue is that which discharges a rarified light, as in bodies which consist of white and black particles arranged alternately. Green is nearly allied to blue. Yellow is a mixture of much white and a little red; and Purple consists of a great deal of blue mixed with a small portion of red. The blue colour of the sea arises from the whiteness of the salt which it contains, mixed with the blackness of the pure water in which the salt is dissolved; and the blueness of the shadows of bodies, seen at the same time by candle and daylight, arises from the whiteness of the paper mixed with the faint light or blackness of the twilight". [19] Here we find that the Grosseteste's framework of the combinations of the features of light (copious and clear) and of transparent media (pure or not) is maintained in the Barrow's approach to colours.

The first Newton's studies on prisms were made on 1666, aiming to improve the optical instruments, in particular telescopes. He found the white light a mixture of colours refracted differently by a transparent medium. Experimenting with two prisms, he showed that a second prism can be used to put back together the light into white light [19]. If the origin of colours were the impurity of the transparent medium, this recombination would be impossible. Newton therefore, with his experiments, disrupted the medieval framework of optics, showing that it is not the purity or impurity of a medium that, interacting with light, gives the colours, but the different refractions of the components of the white light.

**References**

[1] A.C. Sparavigna, Translation and discussion of the De Iride, a treatise on optics by Robert Grosseteste, arXiv, 2012, History and Philosophy of Physics, arXiv:1211.5961, http://arxiv.org/abs/1211.5961

[2] Lewis, Neil, Robert Grosseteste, The Stanford Encyclopedia of Philosophy (Winter 2010 Edition), http://plato.stanford.edu/entries/grosseteste/

[3] These six features are purity and impurity for the medium and brightness, obscurity, intensity and tenuity for light.

[4] D. Harper, Online Etymology Dictionary, 2012 http://www.etymonline.com/index.php

[5] Pliny the Elder, The Natural History, by John Bostock, London. Taylor and Francis, Red Lion Court, Fleet Street, 1855.

[6] The Latin text is that given y "The Electronic Grosseteste", http://www.grosseteste.com/, which is reporting the printed Source: Die Philosophischen Werke des Robert Grosseteste, Bischofs von Lincoln, W. Aschendorff, 1912.

[7] A.C. Sparavigna, Sound and motion in the De Generatione Sonorum, a treatise by Robert Grosseteste, arXiv, 2012, History and Philosophy of Physics, arXiv:1212.1007, http://arxiv.org/abs/1212.1007

[8] A.C. Sparavigna, Discussion of the De Generatione Sonorum, a treatise on sound and phonetics by Robert Grosseteste, Scribd, December 5, 2012, available at http://www.scribd.com/doc/115581526/Discussion-of-the-De-Generatione-Sonorum-a-treatise-on-sound-and-phonetics-by-Robert-Grosseteste

[9] A.C. Sparavigna, The play of colours of prisms, arXiv, 2012, Popular Physics, arXiv: 1207.3504, http://arxiv.org/abs/1207.3504
[10] Herbert Grabes, The Mutable Glass: Mirror Imagery in Titles and Texts of the Middle Ages and the English Renaissance, Cambridge University Press, 1982, pag.43.
[11] Henricus Bate, Speculum divinorum et quorundam naturalium, parts XIII-XVI, edited by Guy Guildentops, Leuven University Press, 2002.
[12] http://en.wikipedia.org/wiki/Black#The_Middle_Ages
[13] Michel Pastoureau, Noir: Histoire d'une couleur, Paris, Seuil, 2008
[14] Helmut Gebelein, Alchimia, La maiga della sostanza, 2009, Edizioni Mediterranee, Roma.
[15] Narciso Silvestrini and Ernst Peter Fischer, Colorsystem, Colour order systems in art and science, at the webpage http://www.colorsystem.com/?page_id=551&lang=en
[16] H.E. Smithson, G. Dinkova-Bruun, G.E.M. Gasper, M. Huxtable, T.C.B. McLeish, and C. Panti, A three-dimensional color space from the 13$^{th}$ century, J. Opt. Soc. Am. Opt. Image Sci. Vis. 2012 Ferbuary 1, 29(2): A346-A352.
[17] Clare C. Riedl, Robert Grosseteste On light, Marquette University Press, Milwaukee, Wisconsin, 1942.
[18] D. Allchin, Newton's Colors, SHiPS Resource Center. Retrieved 2010-10-16.
[19] David Brewster, The life of Sir Isaac Newton, Harper & Brothers, New York, 1840.

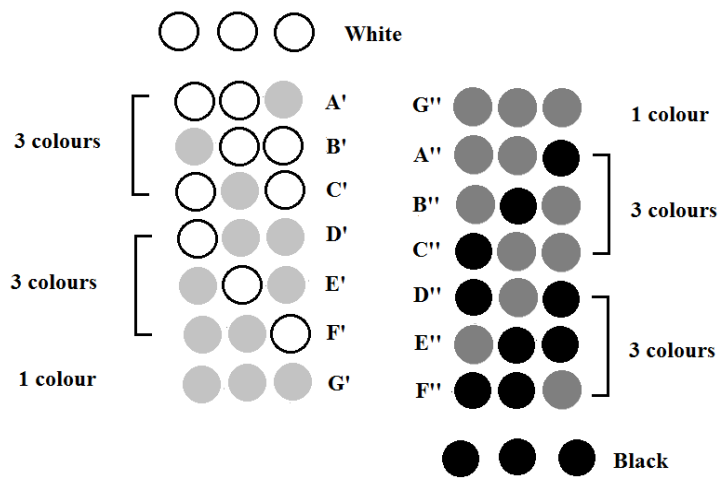

Figure 1: The colours can be created by white, which is brightness, intensity and purity. We can relax one, two or all these three features to obtain seven colours, as in the calculus of combinations. The relaxation of one of the white features is rendered by a grey circles. The same we can do from the black.

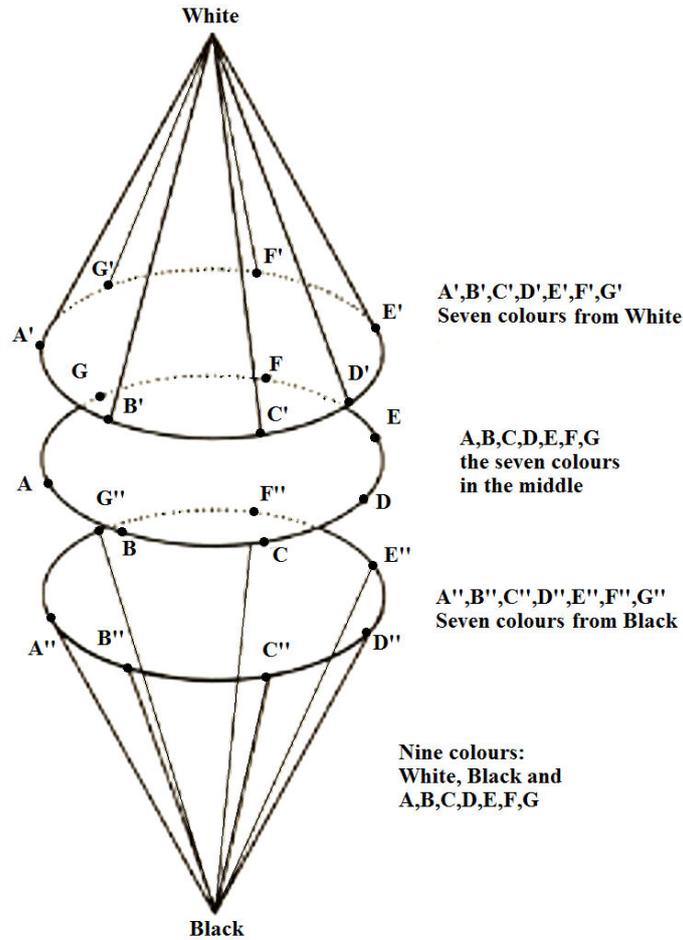

Figure 2: The seven colours from white and the seven colours from black are merging in "average" colours, which gives a circle of seven colours [15]. White, black and these seven colours provide nine colours.

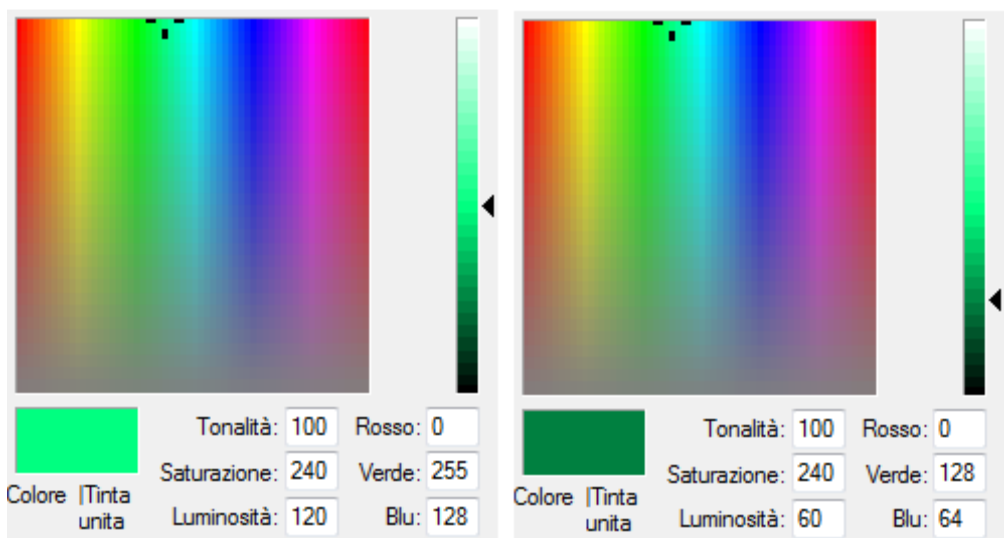

Figure 3 – An example from the palette of Paint software. We see a light green and a dark green. In any case, it is green.